\documentclass[aps,prl,reprint,superscriptaddress,showpacs]{revtex4-1}
\usepackage{graphicx}

\usepackage{color}
\usepackage{ulem}

\begin{document}

\title{Composition-induced structural instability and strong-coupling superconductivity in Au$_{1-x}$Pd$_x$Te$_2$}

\author{Kazutaka Kudo}
\email{kudo@science.okayama-u.ac.jp}
\affiliation{Department of Physics, Okayama University, Okayama 700-8530, Japan}
\affiliation{Research Institute for Interdisciplinary Science, Okayama University, Okayama 700-8530, Japan}\author{Hiroyuki Ishii}
\affiliation{Department of Physics, Okayama University, Okayama 700-8530, Japan}
\author{Minoru Nohara}
\email{nohara@science.okayama-u.ac.jp}
\affiliation{Department of Physics, Okayama University, Okayama 700-8530, Japan}
\affiliation{Research Institute for Interdisciplinary Science, Okayama University, Okayama 700-8530, Japan}

\date{}

\begin{abstract}
The physical properties and structural evolution of the MX$_2$-type solid solution Au$_{1-x}$Pd$_x$Te$_2$ are reported. 
The end member AuTe$_2$ is a normal metal with a monoclinic distorted CdI$_2$-type structure with preformed Te--Te dimers. 
A monoclinic--trigonal structural phase transition at a finite temperature occurs upon Pd substitution and is suppressed to zero temperature near $x$ = 0.55, and a superconducting phase with a maximum $T_{\rm c}$ = 4.65 K emerges. 
A clear indication of strong coupling superconductivity is observed near the composition of the structural instability. 
The competitive relationship between Te--Te dimers and superconductivity is proposed.
\end{abstract}



\maketitle

Superconductivity at a relatively high transition temperature ($T_{\rm c}$) often emerges near structural instability that is characterized by pressure- or composition-induced structural phase transition. 
Typical examples of such superconductivity are of iron and nickel pnictides \cite{Cruz,Niedziela,Goto,Yoshizawa,Kudo1,Hirai}, 
iridium and gold tellurides \cite{Pyon,Yang,Kamitani,Kudo2,Kudo3,Kitagawa}, 
A15 compounds \cite{Testardi}, 
graphite intercalated compounds \cite{Gauzzi,Gauzzi2}, 
and quasiskutterudite stanides \cite{Klintberg,Goh,Yu}. 
Among them, iridium and gold tellurides, namely IrTe$_2$ and AuTe$_2$ with distorted CdI$_2$-type structures, have been attracting considerable interest because their structural instabilities result from the breaking of moleculelike dimers of iridium \cite{Pascut,Toriyama} or tellurium \cite{Kudo3,Kitagawa}, and the subsequent emergence of a superconducting phase upon applying hydrostatic pressure \cite{Kitagawa} or chemical doping \cite{Pyon,Yang,Kamitani,Kudo2,Kudo3}. 
The evolution of electronic states across the structural transition has been intensively studied on IrTe$_2$ \cite{Ootsuki1,Ootsuki2,Takubo,Ko}, while the study of AuTe$_2$ is limited \cite{Ootsuki_AuTe2} because composition-induced structural instability has not yet been exhibited experimentally. 

AuTe$_2$, known as mineral calaverite, crystallizes in a monoclinic distorted CdI$_2$-type structure with the space group $C2/m$ ($C^{3}_{2h}$, No.~12) \cite{Tunell}. 
Each AuTe$_2$ layer consists of edge-shared AuTe$_6$ octahedra that are strongly distorted with two short (2.67 {\AA}) and four long (2.98 {\AA}) Au-Te bonds in the average structure \cite{Tunell}. This is due to the formation of Te--Te dimers with a bond length of 2.88 {\AA} between the layers \cite{Schutte}, which results in an incommensurate modulation of {\bf q} = $-$0.4076{\bf a$^*$}+0.4479{\bf c$^*$}. 
Recently, Kitagawa {\it et al.} demonstrated that AuTe$_2$ exhibits pressure-induced structural instability that is characterized by a monoclinic--trigonal structural phase transition,  together with the subsequent emergence of a superconducting phase with a maximum $T_{\rm c}$ = 2.3 K \cite{Kitagawa}. 
Kudo {\it et al.} reported superconductivity at $T_{\rm c}$ = 4.0 K in the solid solution Au$_{1-x}$Pt$_x$Te$_2$ with $x$ = 0.35 \cite{Kudo3}. 
In both cases, superconductivity emerges in the undistorted trigonal CdI$_2$-type structure with the space group $P\bar{3}m1$ ($D^{3}_{3d}$, No.~164) \cite{Reithmayer,Kitagawa,Kudo3}, where Te--Te dimers are broken. 
However, the pronounced phase separation that occurs at 1.6 $<$ $P$ $<$ 2.7 GPa in AuTe$_2$ under pressure \cite{Kitagawa} or 0.1 $<$ $x$ $<$ 0.15 in the solid solution Au$_{1-x}$Pt$_x$Te$_2$ \cite{Kudo3} has inhibited us from accessing the critical region of structural instability. 
Thus, we should search for another doping element that forms continuous solid solution in AuTe$_2$. 

In this paper, we report on the physical properties and structural evolution of Au$_{1-x}$Pd$_x$Te$_2$, which forms continuous solid solution across structural instability. 
We demonstrate the systematic suppression of the monoclinic Te--Te dimer phase of AuTe$_2$ by Pd substitution. 
A superconducting phase emerges when the monoclinic phase is suppressed and the trigonal phase appears at $x$ = 0.55. 
The specific heat and magnetization data suggest that the enhanced electronic density of states (DOS) at the Fermi level $E_{\rm F}$ is responsible for the observed strong-coupling superconductivity. 
On the other hand, the DOS at $E_{\rm F}$ is strongly suppressed in the monoclinic phase, suggesting the competition between Te--Te dimers and superconductivity. 
Our finding demonstrates that the breaking of moleculelike dimers in solids offers a novel route to develop superconductors.

Polycrystalline samples of Au$_{1-x}$Pd$_x$Te$_2$ with nominal compositions of 0.00 $\leq$ $x$ $\leq$ 1.00 were synthesized using a solid-state reaction. 
First, stoichiometric amounts of Au (99.99\%), Pd (99.98\%), and Te (99.99\%) were mixed and pulverized. 
They were heated at 500$^\circ$C for 24 h in an evacuated quartz tube. 
Subsequently, the product was powdered, pressed into pellets, and annealed at 350--700$^\circ$C for 24 h in an evacuated quartz tube.
The annealing was performed once or twice to homogenize the sample. 
The heating and cooling rates both equaled 20$^\circ$C/h. 
The resulting samples were characterized at room temperature by powder x-ray diffraction (XRD) using a Rigaku RINT-TTR III x-ray diffractometer with Cu $K\alpha$ radiation and were identified to be a single phase of Au$_{1-x}$Pd$_x$Te$_2$ \cite{cubic_alloy}. 
Energy dispersive x-ray spectrometry (EDS) was used to determine the $x$. 
The measured $x$ values were in good agreement with the nominal ones; we used the nominal $x$ in this study. 
Magnetization $M$ was measured using a Quantum Design magnetic property measurement system (MPMS). 
Electrical resistivity $\rho$ and specific heat $C$ were measured using a Quantum Design physical property measurement system (PPMS).

\begin{figure}[t]
\begin{center}
\includegraphics[width=6.0cm]{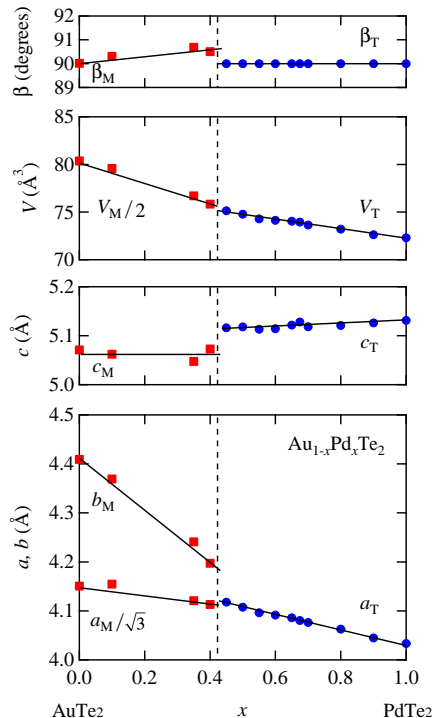}
\caption{
Room-temperature lattice parameters for Au$_{1-x}$Pd$_x$Te$_2$. Subscripts M and T respectively indicate monoclinic and trigonal phases. 
}
\end{center}
\end{figure}
\begin{figure}[t]
\begin{center}
\includegraphics[width=7cm]{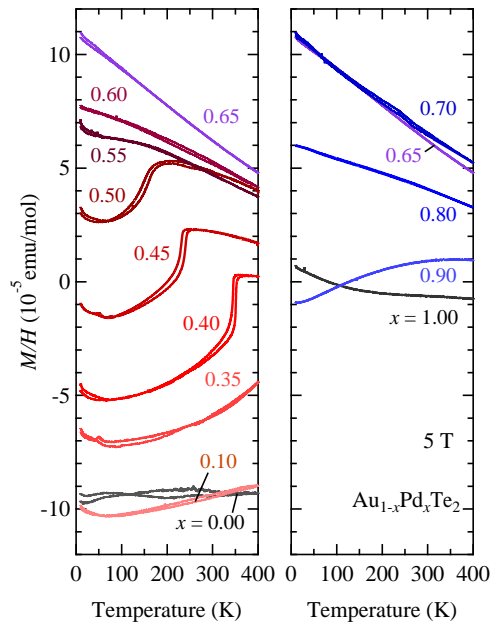}
\caption{
Temperature dependence of magnetization divided by magnetic field, $M/H$, in the magnetic field of 5 T for Au$_{1-x}$Pd$_x$Te$_2$. Data were measured upon heating and cooling. The core diamagnetism for Pd, Au, and Te has not been corrected.}
\end{center}
\end{figure}
\begin{figure}[t]
\begin{center}
\includegraphics[width=6cm]{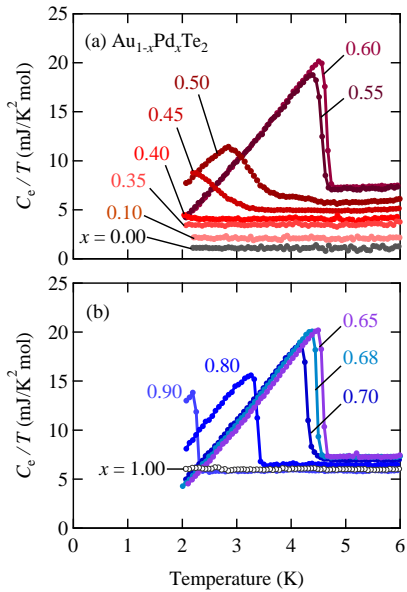}
\caption{
Temperature dependence of electronic specific heat divided by temperature, $C_{\rm e}/T$, for Au$_{1-x}$Pd$_x$Te$_2$, in which $C_{\rm e}$ is the difference of total specific heat $C$ and phonon contribution. 
}
\end{center}
\end{figure}

The structural instability that results from Te--Te dimer breaking, which can be recognized as the structural transition from a monoclinic to trigonal phase, was observed at $x$ = 0.4 at room temperature, as shown in Fig.~1. 
The XRD profiles for $x \le 0.40$ can be indexed based on the monoclinic average structure of end-member AuTe$_2$; as the Pd content increases, the parameters $a$ and $b$ decrease, while the parameter $c$ shows no substantial change. 
Between $x =$ 0.40 and 0.45, the structural phase transition to a trigonal phase occurs, indicating the breaking of Te--Te dimers by Pd doping. 
The discontinuous changes in the lattice parameters suggest a first-order phase transition. 
The parameter $a$ slightly increases, the parameter $b$ decreases, and the parameter $c$ increases. 
Therefore, the resulting change in cell volume $V$ is small.

The structural transition also depends on temperature. 
As is shown in Fig.~2, the temperature dependence of magnetization shows drops at 350 K for $x =$ 0.40 and 240 K for $x$ = 0.45, respectively, suggesting the reduction of the DOS  at $E_{\rm F}$. The drops in $M/H$ is one order of magnitude smaller than those in systems that exhibit a metal--insulator transition \cite{Katayama}. 
In response to this, the electrical resistivity exhibits a jump at the same temperature, but it remains metallic at low temperatures (see Supplemental Material A \cite{Supplement}). 
The anomalies can be ascribable to the trigonal-to-monoclinic phase transition resulted from the formation of Te--Te dimers upon cooling, because the samples of $x =$ 0.40 and 0.45 are identified as the monoclinic and trigonal phases, respectively, at room temperature, as shown in Fig.~1. 
The observed thermal hysteresis in the temperature dependence of magnetization and resistivity is consistent with the first-order phase transition, which is implied by the $x$ dependence of the lattice parameters. 
We determined the structural phase transition temperature ($T_{\rm s}$) from the drops in magnetization and jumps in resistivity. 
The $T_{\rm s}$ decreases with increasing Pd content and becomes absent at $x \ge$ 0.55, suggesting that the isolated Te is stabilized down to $T =$ 0 K in the compositions. 

\begin{figure}[t]
\begin{center}
\includegraphics[width=6cm]{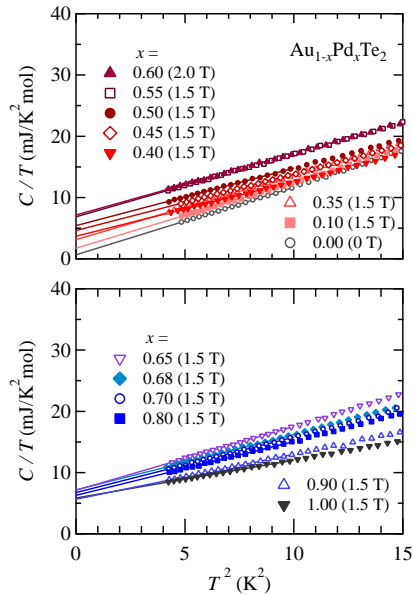}
\caption{
The specific heat divided by temperature, $C/T$, as a function of $T^2$ for Au$_{1-x}$Pd$_x$Te$_2$. 
Solid lines denote fitted equation $C/T = \gamma + \beta T^2$, where $\gamma$ is the electronic specific-heat coefficient and $\beta$ is a constant that corresponds to the Debye phonon contribution. 
}
\end{center}
\end{figure}

\begin{figure}[t]
\begin{center}
\includegraphics[width=7cm]{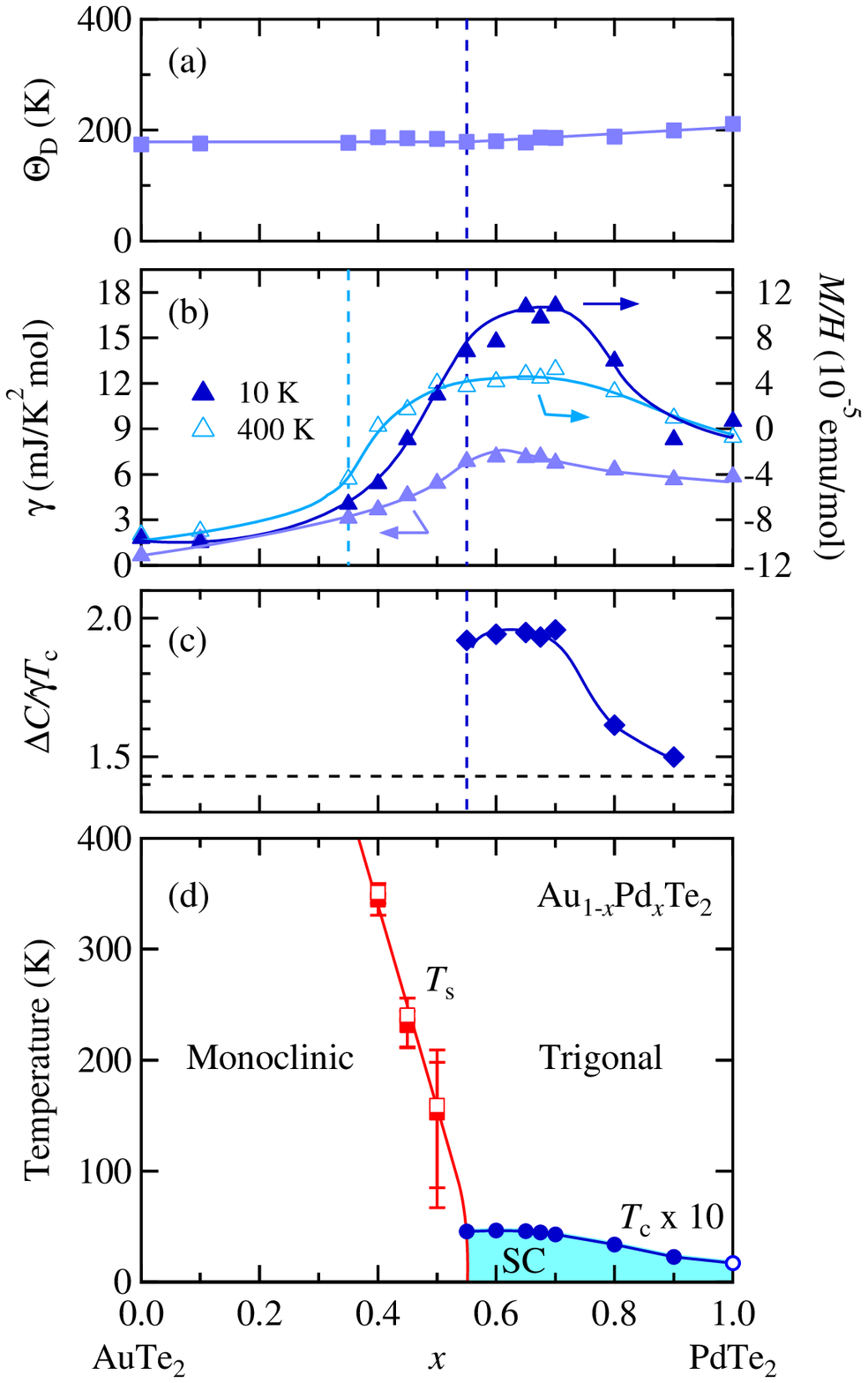}
\caption{
(a) Debye temperature $\Theta_{\rm D}$, (b) electronic specific-heat coefficient $\gamma$ and magnetic susceptibility $M/H$ with temperature 10 and 400 K, and (c) normalized specific heat jump $\Delta C / \gamma T_{\rm c}$ at the superconducting transition as a function of $x$ for Au$_{1-x}$Pd$_x$Te$_2$, in which the horizontal dotted line corresponds to a BCS weak-coupling value of $\Delta C / \gamma T_{\rm c} = 1.43$. 
(d) Electronic phase diagram of Au$_{1-x}$Pd$_x$Te$_2$, in which the (blue) closed circles represent the superconducting transition temperatures, $T_{\rm c}$ for $0.55 \le x \le 0.90$, that were determined from the specific heat measurements, and the (blue) open circle indicates $T_{\rm c}$ for $x =$ 1.00 provided by Ref.~\onlinecite{Raub}. 
SC denotes the superconducting phase, and the (red) closed and open squares represent the trigonal-to-monoclinic structural phase transition temperatures, $T_{\rm s}$, determined from the magnetization measurements upon cooling and heating, respectively. 
The solid curves are guides. 
}
\end{center}
\end{figure}

Along with the disappearance of Te--Te dimers, a superconducting phase appears.
As shown in Fig.~3, the clear jump of the electronic specific heat ($C_{\rm e}$) indicates the emergence of bulk superconductivity for $x \geq 0.55$, while the smeared jumps at $x =$ 0.45 and 0.50 indicate the absence of bulk superconductivity in these samples. 
The maximum $T_{\rm c}$ of 4.65 K is observed at $x =$ 0.60, and further Pd doping lowers $T_{\rm c}$ towards 1.69 K of $x$ = 1.00 \cite{Raub}. 
The normalized specific-heat-jump at the superconducting transition ($\Delta C / \gamma T_{\rm c}$) is 1.50 for $x =$ 0.90 which agrees with the Bardeen-Cooper-Schrieffer (BCS) weak-coupling value of 1.43, whereas $\Delta C / \gamma T_{\rm c} \simeq$ 1.94 for $x \sim$ 0.60 which corresponds to the value of strong electron-phonon coupling superconductors \cite{Carbotte}. 
The superconducting transitions were also demonstrated by the sharp resistivity transition and the full shielding diamagnetic signal (see Supplemental Material B \cite{Supplement}).

The strong-coupling superconductivity observed in Pd-doped AuTe$_2$ is attributed to the enhanced DOS at $E_{\rm F}$. 
Theoretically, either DOS enhancement or phonon softening can increase electron-phonon coupling \cite{McMillan}. 
However, a standard analysis of the low-temperature specific-heat data indicates that the phonon softening in the material is very small. 
The normal-state heat capacity data under an applied field that suppresses superconductivity is well fitted by equation $C/T = \gamma + \beta T^2$, as shown in Fig.~4, where $\gamma$ is the electronic specific-heat coefficient and $\beta$ is the phonon contribution.  
According to Fig.~5(a), estimated Debye temperature $\Theta_{\rm D}$ exhibits little change ($<$ 5\%) as a function of $x$, even though the system approaches the structural phase boundary. 
On the other hand, $\gamma$ increases with decreasing Pd content in the trigonal side and achieves a maximum at the $x$ where $T_{\rm c}$ exhibits the maximum value, as shown in Fig.~5(b). 
This is consistent with the magnetization data; the $M/H$ of PdTe$_2$ ($x$ = 1.00) is almost zero and the value rapidly increases with decreasing Pd content in the trigonal side, as shown in Figs.~2 and 5(b). 
Thus, the magnetization and specific-heat results indicate that the strong-coupling superconductivity in the present system is exclusively attributed to the electronic origin. 
This highly contrasts with BaNi$_2$As$_2$, in which strong-coupling superconductivity is accompanied by a drastic phonon softening ($>30$\%) with no visible enhancement in the DOS \cite{Kudo1}.

Our results are summarized in Fig.~5.
The monoclinic Te--Te dimer phase in AuTe$_2$ is suppressed by Pd doping, and varnishes at $x_{\rm c} =$ 0.55, as shown in Fig.~5(d). 
As soon as the Te--Te dimers disappear, a superconducting phase emerges in the trigonal phase, suggesting the competitive relationship between Te--Te dimers and superconductivity. 
This competition is ascribable to the reduction of DOS at $E_{\rm F}$, because the $\gamma$ and $M/H$ in the monoclinic phase are strongly suppressed, as shown in Fig.~5(b).
On the other hand, the strong-coupling superconductivity results from the enhanced DOS in the trigonal phase, as shown in Figs.~5(b) and 5(c). 
Both $\gamma$ and $M/H$ exhibit a broad maximum at $x_{\rm m}$ $\simeq$ 0.65, which is noticeably apart from the monoclinic--trigonal phase boundary. 
The maximum is prominent in the $x$-dependent $M/H$ at 10 K and $x$ $\geq$ 0.55 and even at 400 K and $x$ $\geq$ 0.40 in the trigonal phase, as shown in Fig.~5(b). 
Here, we note that the Wilson ratio $\Delta \chi / \gamma$ (in units of $3\mu_{\rm B}^{2}/\pi{^2}k_{\rm B}^{2}$) at $x_{\rm m}$, in which $\Delta \chi$ corresponds to the difference in $M/H$ between $x = 0.00$ and $x_{\rm}$ = 0.65 at 10 K, is estimated to be 2.0 \cite{Wilson}. 
This value could suggest the enhanced electronic correlation around $x_{\rm m}$ (see Supplemental Material C \cite{Supplement}). 
The remarkable increase in $M/H$ caused by lowering temperature as well as the unusual $T$-linear temperature dependence in $M/H$ for $x$ = 0.65, shown in Fig. 2, might also suggest it. 
To address this issue, detailed investigation is expected with consideration for a possible proximity to a van Hove singularity, which is associated with the doping-dependent DOS maximum \cite{Kitagawa,Myron,Jan}, as well as the structural instability that results from Te--Te dimer breaking.

In conclusion, our experiments demonstrate the emergence of strong-coupling superconductivity, which is associated with the enhancement of the electronic density of states in palladium-doped AuTe$_2$. 
The superconductivity sets in as soon as the breaking of Te--Te dimers. 
The revealed competition between Te--Te dimers and superconductivity in the present system suggests that dimer breaking would invoke novel superconductivity in a wide variety of materials.

Some of this research was performed at the Advanced Science Research Center, Okayama University. 
This work was partially supported by Grants-in-Aid for Scientific Research (Grants No. 25400372,  No. 26287082, No. 15H01047, and No. 15H05886) provided by the Japan Society for the Promotion of Science (JSPS).


\begin{thebibliography}{99}
\bibitem{cubic_alloy} The present syntheses did not provide metastable simple-cubic alloys Au-Te \cite{Newkirk} and Au-Pd-Te \cite{Chen}, which demonstrate superconductivity. However, this is reasonable because the simple-cubic phase is never obtained unless rapid quenching is performed from the liquid phase at a rate of approximately 10$^7$$^\circ$C/s \cite{Chen}.
\bibitem{Wilson} Here, we assumed that the value of $\chi$ ($=M/H$) for AuTe$_2$ is dominated by core diamagnetism because of the substantially reduced $\gamma$ value in the monoclinic phase, and thus $\Delta \chi$ approximately represents Pauli paramagnetic susceptibility.
\end{thebibliography}
%

\clearpage

\newcommand{\beginsupplement}{%
        \setcounter{table}{0}
        \renewcommand{\thetable}{S\arabic{table}}%
        \setcounter{figure}{0}
        \renewcommand{\thefigure}{S\arabic{figure}}%
     }
     
\beginsupplement

\section*{Supplemental materials}

\subsection{A. Structural phase transition}

\begin{figure}[h]
\begin{center}
\includegraphics[width=5.5cm]{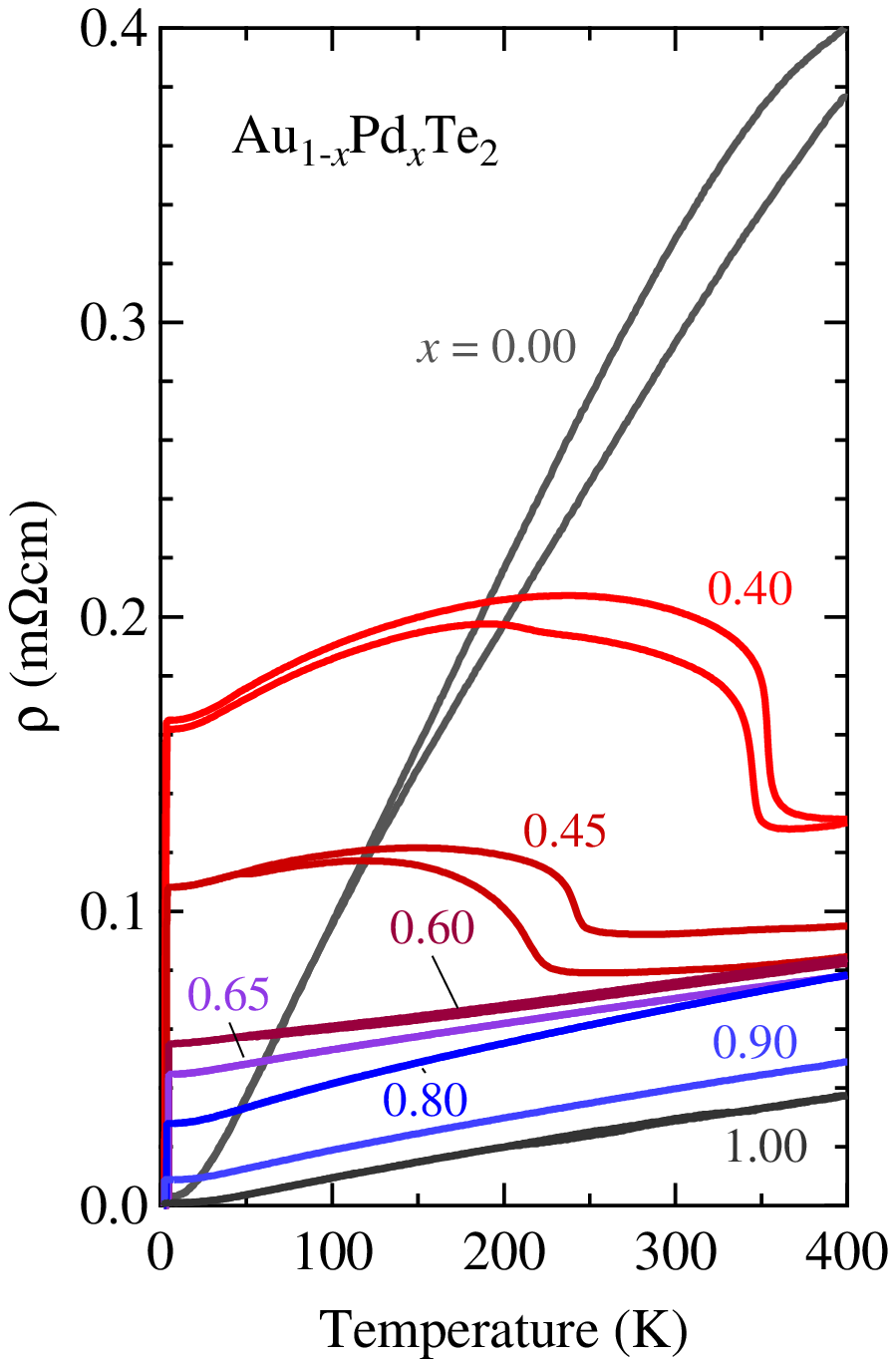}
\caption{
Temperature dependence of electrical resistivity $\rho$ for Au$_{1-x}$Pd$_x$Te$_2$.
}\label{FigA}
\end{center}
\end{figure}

Figure \ref{FigA} shows the electrical resistivity $\rho$ for Au$_{1-x}$Pd$_x$Te$_2$. 
$\rho$ shows a clear jump at the structural phase transition in the samples with $x$ = 0.40 and 0.45. 
The transition temperatures agree well with those determined from the magnetization. 
Consistent with the magnetization results, the transition temperature at which a jump is observed decreases upon Pd doping. 
A similar resistive anomaly was also reported in the high-pressure study of AuTe$_2$ \cite{Kitagawa}. 

\subsection{B. Superconductivity}

\begin{figure}[h]
\begin{center}
\includegraphics[width=9cm]{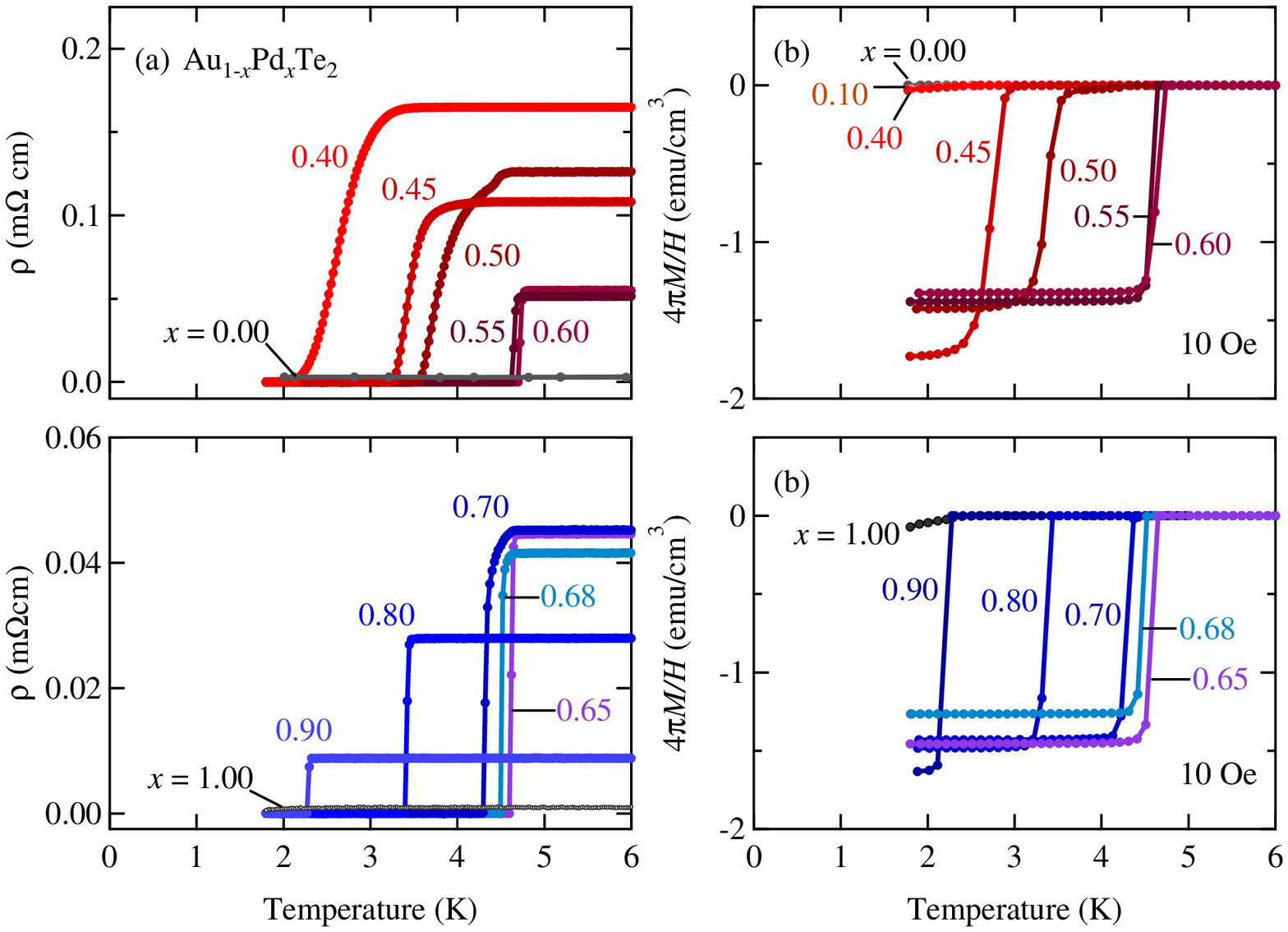}
\caption{
(a) Temperature dependence of electrical resistivity $\rho$ for Au$_{1-x}$Pd$_x$Te$_2$. 
(b) Temperature dependence of magnetization divided by magnetic field, $M/H$, in a magnetic field with 10 Oe for Au$_{1-x}$Pd$_x$Te$_2$ under zero-field cooling, in which no correction for the diamagnetizing field has been made. 
}\label{FigB}
\end{center}
\end{figure}

Figure \ref{FigB} shows a dependence of electrical resistivity $\rho$ and magnetization $M$ on temperature for Au$_{1-x}$Pd$_x$Te$_2$ at low temperatures. 
The superconducting transition temperatures $T_{\rm c}$ determined from $\rho$ and $M$ are consistent with that of specific heat $C$. 

\subsection{C. Wilson ratio}

\begin{figure}[h]
\begin{center}
\includegraphics[width=6cm]{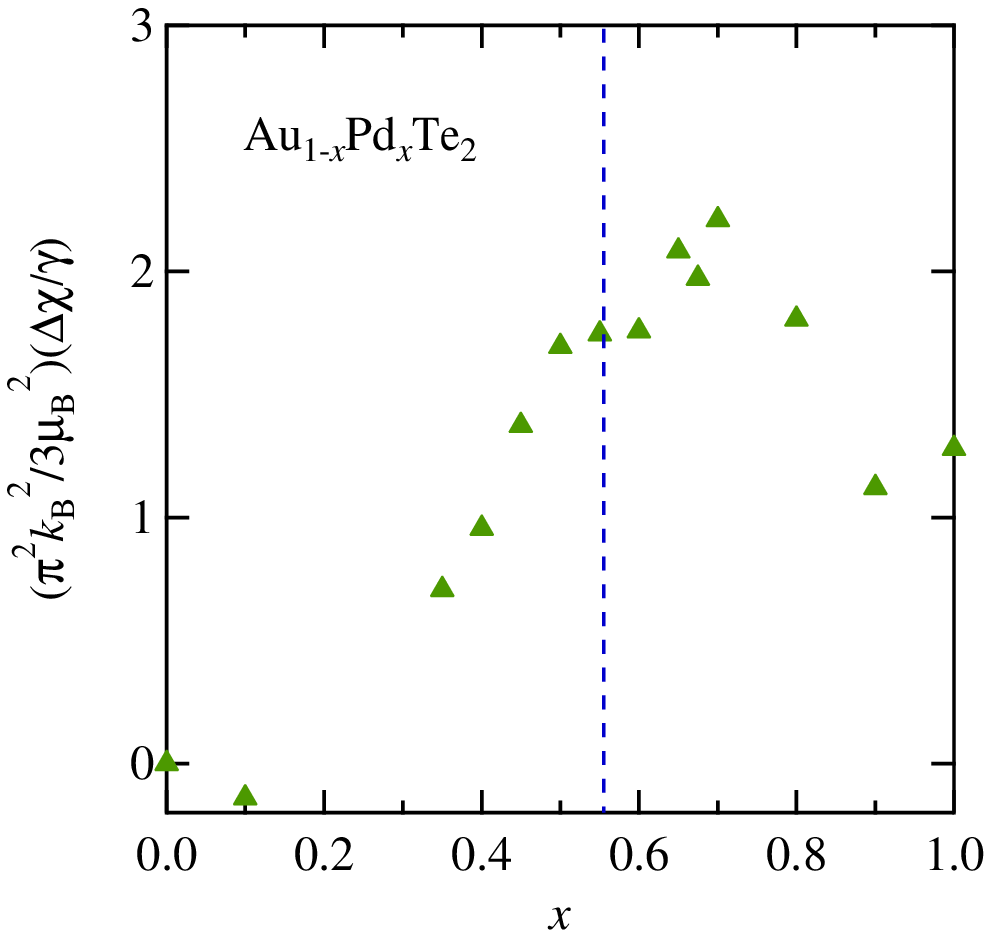}
\caption{
Doping dependent Wilson ratio for Au$_{1-x}$Pd$_x$Te$_2$. 
}\label{FigC}
\end{center}
\end{figure}

Figure \ref{FigC} shows a dependence of the Wilson ratio $\Delta \chi / \gamma$ (in units of $3\mu_{\rm B}^{2}/\pi{^2}k_{\rm B}^{2}$) on palladium content $x$ for Au$_{1-x}$Pd$_x$Te$_2$. 
Here, $\Delta \chi$ corresponds to the difference in $M/H$ between $x = 0.00$ (AuTe$_2$) and $x \neq 0.00$ (Au$_{1-x}$Pd$_x$Te$_2$) at 10 K. 
We assumed that the value of $\chi$ ($=M/H$) for AuTe$_2$ is dominated by core diamagnetism because of the substantially reduced $\gamma$ value in the monoclinic phase, and thus $\Delta \chi$ approximately represents Pauli paramagnetic susceptibility.

\end{document}